\documentclass[letterpaper]{article}

\usepackage{natbib,alifeconf}
\usepackage{hyperref}

\title{Social learning in a simple task allocation game}
\author{Rui Chen$^{1}$, Julian Garc\'ia$^{1}$ \and Bernd Meyer$^1$ \\
\mbox{}\\
$^1$Faculty of Information Technology \\ Monash University \\
Australia\\
\{rui.chen, julian.garcia, bernd.meyer\}@monash.edu}

\begin{document}
\maketitle

\section{Background}
The ability of social insect colonies to tackle different tasks simultaneously without central control is a key factor in their ecological success. Individual choices by colony members give rise to collective behaviours that are robust to dynamic and uncertain environments~\citep{cdfstb-2001}. The mechanisms of this process, known as self-organised task allocation, have inspired a variety of applications in Computer Science, including scheduling~\citep{cbtd-2000} and control~\citep{DresslerSO}.

Individual choices in task allocation are determined by three different interacting factors: individual personalities, the environment, and interactions amongst colony members~\citep{Charbonneau:2015ew}. The interplay between individual traits and the environment has been studied extensively using  models~\citep{btd-1996,Duarte:2012cw,Lichocki:2012tk}; but the effect of  social interactions is less well-understood ~\citep{Jeanson:2013bs, Kao2014}. 

We investigate the effects of social interactions in task allocation using Evolutionary Game Theory (EGT).  We propose a simple task-allocation game and study how different learning mechanisms can give rise to specialised and non-specialised colonies under different ecological conditions. By combining agent-based simulations and adaptive dynamics we show that social learning can result in colonies of generalists or specialists, depending on ecological parameters. Agent-based simulations further show that learning dynamics play a crucial role in task allocation. In particular, introspective individual learning readily favours the emergence of specialists, while a process resembling task recruitment favours the emergence of generalists.     

\section{A Simple Task-Allocation Game}

Our model assumes that individuals live in groups of size $n$, and choose to allocate their effort into two different tasks: Task A is a regulatory task, such as fanning to cool down nest temperature; Task B is a foraging task, required to support the energy costs of the colony.  The behaviour of individual  $i$ is determined by a continuous trait $x_i \in \left[0 ,1\right]$, the probability that individual $i$ will perform Task A. Thus, $1-x_i$ is the probability of individual $i$ engaging in Task B. In the most general sense, the payoff of individual $i$ is given by: $\Pi(x_i) = B(x_i) - C(x_i)$, where $B(x_i)$ and $C(x_i)$ are benefits and cost respectively. Total payoff for one individual depends not only on her own trait, but also on the traits of all others in the population.

We define the benefit function as $B(x_i) = B^A(x_i) \cdot B^B(x_i)$. This implies that the benefits of foraging are discounted if the colony is not well-regulated. For the regulatory task, we assume that benefits are maximised when an intermediate number of individuals are engaged in regulation, thus $B^A(x_i)$ is a concave function in $\sum_{j=0}^{n} x_j$. For the foraging task we assume that individual benefits increase linearly with collective foraging effort.

To fix ideas $B^A(x_i)$ is assumed to be quadratic with maximum value 1 when half of the workers in the group are engaged in Task A. Likewise, $b$ is the individual reward from performing Task B in one time period. Thus, 
$$B^A(x_i) = -\frac{4}{n^2}\cdot (\sum_{j=1}^{n}x_j)^2 + \frac{4}{n}\cdot \sum_{j=1}^{n}x_j$$
and
$$B^B(x_i) = \frac{1}{n}\cdot b\bigg(\sum_{j=1}^{n}(1-x_j)\bigg).$$

For this payoff structure the rewards of foraging are maintained in full, only when regulation is optimal. 

Costs are defined as $C(x_i) = C^A(x_i) + C^B(x_i)$. The cost of regulation is assumed to be linear, with $C^A(x_i) = r\cdot x_i$. Here,  $r$ represents the cost of performing Task A for an individual in one time period. The nature of foraging implies decreasing marginal costs, thus  $C^B(x_i) = -(1-x_i)^2 + 2(1-x_i)$ for $x_i \in \left[0, 1 \right]$. 

\section{Social learning}

How do individuals in the colony learn to coordinate their efforts to successfully perform both tasks? We obtain a first approximation by using the mathematical framework of adaptive dynamics~\citep{Brannstrom:2013iu}. The underlying process resembles social learning, whereby individuals tend to copy those that are most successful. The assumptions of adaptive dynamics imply a very large population where mutations are rare and small \citep{waxman200520}. 

The analysis predicts a unique singular strategy $x^*$ that is convergent stable. We derive conditions for evolutionary branching; a situation in which a monomorphous population splits in two morphs \citep{doebeli2004evolutionary}. In the context of task allocation, this implies that the colony successfully tends to both tasks, but each individual specializes by putting all their effort into one single task. Alternatively, the colony can converge to a state in which generalists share responsibilities and each individual splits efforts in both tasks.  We refer to the latter as \emph{weak specialisation}, as opposed to \emph{strong specialisation} in the former case. Given $x^*$, the colony exhibits strong specialisation  if 
$$\frac{8b}{n^2}(3x^*-2) + 2 > 0$$ 
or weak specialisation (evolutionary stable strategy) if 
$$\frac{8b}{n^2}(3x^*-2) + 2 < 0.$$ 
The system can also converge to a state in which the colony fails to coordinate its efforts in both tasks. In this case we speak of an inviable colony. 

\begin{figure}[ht]
	\centering
	\includegraphics[width=0.5\textwidth]{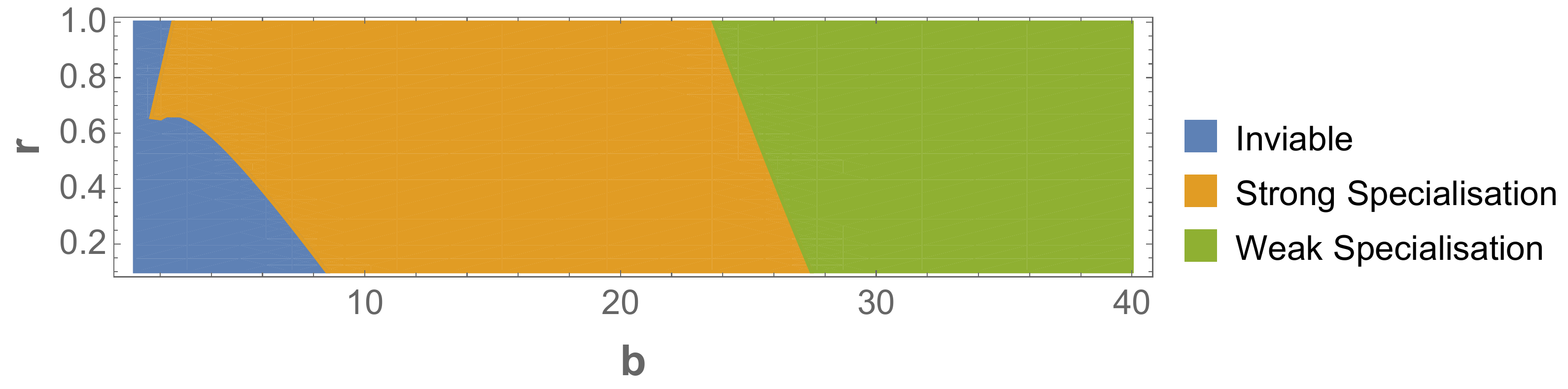}
    \caption{\label{fig:adaptive_dynamics} Classification of colony behaviour as predicted by adaptive dynamics}
\end{figure}

Figure~\ref{fig:adaptive_dynamics} shows how these different outcomes depend on ecological conditions as given by the values of $b$ and $r$.

\begin{figure}[ht]
	\centering
	\includegraphics[width=0.5\textwidth]{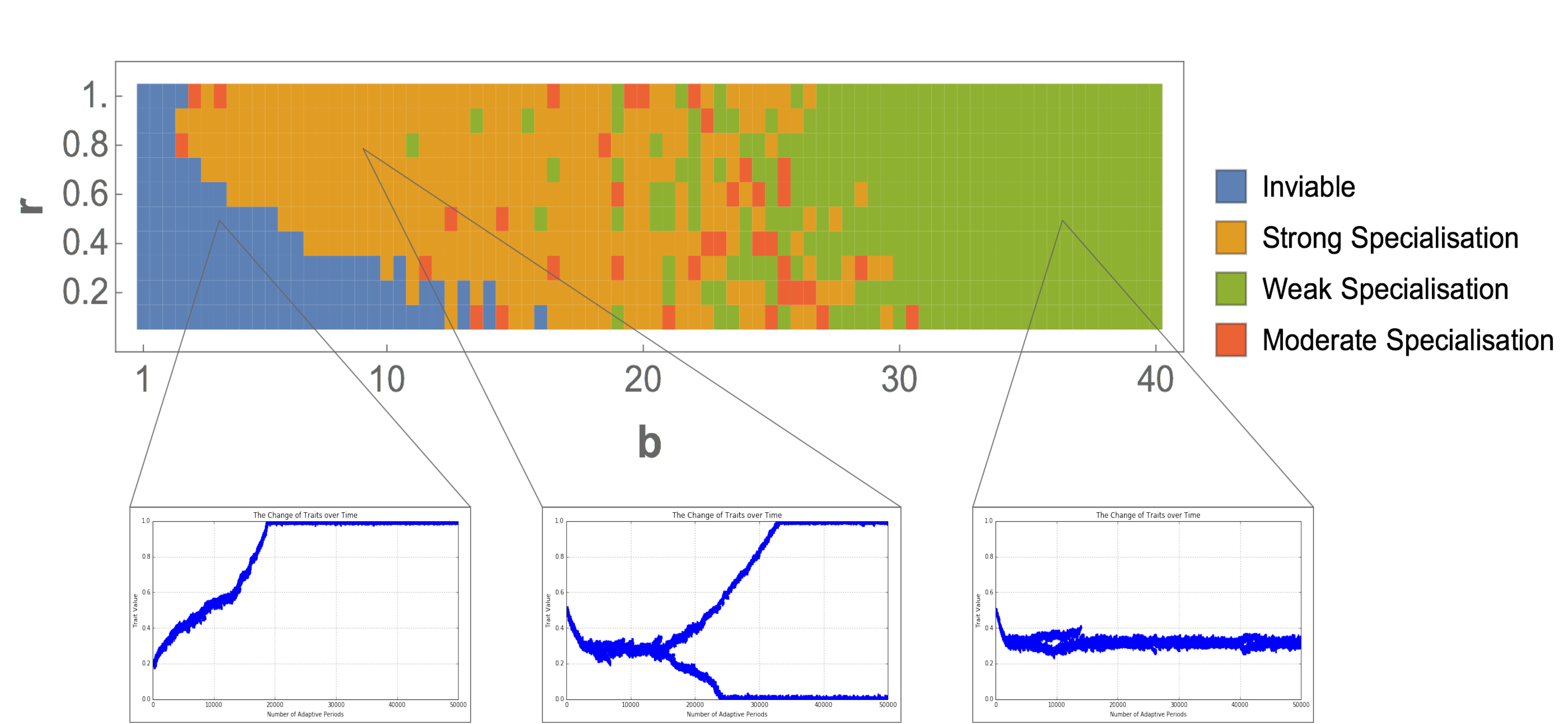}
    \caption{\label{fig:inset_social} Agent-based simulations for social learning}
\end{figure}

To check the robustness of our mathematical prediction we use an agent-based simulation. Social learning follows a Wright-Fisher process where the population is not necessarily monomorphous \citep{imhof2006evolutionary}. This is similar to roulette wheel selection in a standard Genetic Algorithm \citep{fogel2006evolutionary}. 

Figure~\ref{fig:inset_social} shows that our simulation results are in line with the predictions from adaptive dynamics. Our results indicate that when foraging is expensive, agents fail to coordinate and the colony is not viable. For intermediate costs of foraging individuals strongly specialize, while cheap foraging leads to a population of generalists.

\section{Other Learning Mechanisms }

The mathematical framework and the simulations so far assume that agents learn socially. This implies that individuals can observe the traits of all others in the colony, while at the same time inferring how successful a particular trait is. We perform simulations in which this assumption is relaxed in two different ways: \emph{i)} A model of individual learning assumes that agents learn exclusively from their own experience. \emph{ii)} A model of task recruitment assumes that agents respond to recruitment signals for either task. 

\subsection{Individual Learning}

For individual learning we assume that agents remember their most recent strategy and its associated payoff. Innovations arise via exploration with a small probability. If a new strategy is not better than the most recent one, individuals roll back to their previous strategy. This class of introspective learning is thought to be less cognitively demanding than social learning.

\begin{figure}[ht]
	\centering
	\includegraphics[width=\linewidth]{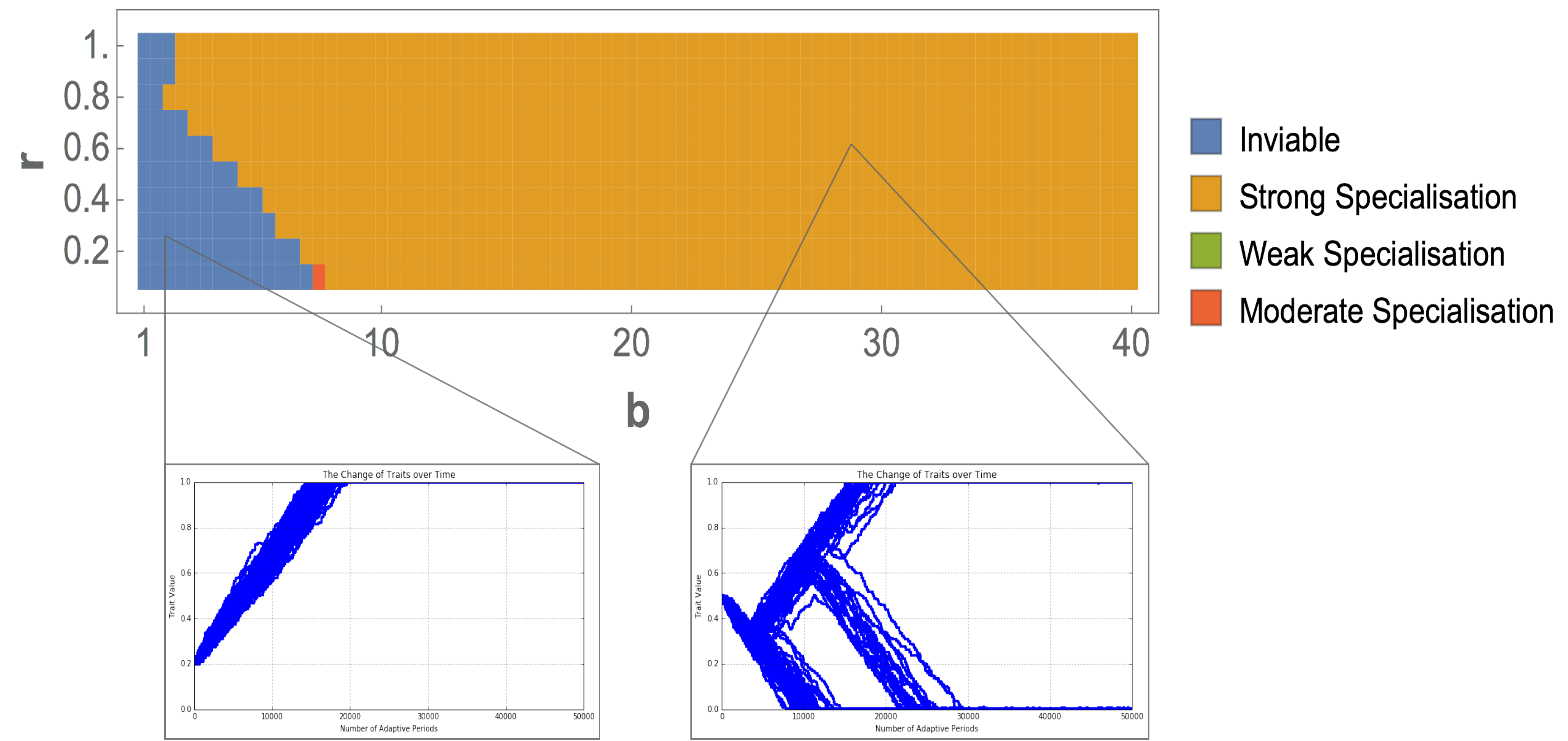}
    \caption{\label{fig:inset_individual} Classification of colonies based on the model of individual reinforcement}
\end{figure}

Figure~\ref{fig:inset_individual} shows the simulation results for this model. Individual learning readily favours strong specialisation across a large range of parameters.

\subsection{Task Recruitment}

We also investigate a process of task recruitment, inspired by the ecological literature on task allocation \citep{Dukas:2008us}. In this version of the model, agents respond to recruitment signals for each task. At each time-step, individuals choose a single task to perform based on their trait, i.e., individual $i$ chooses Task A with probability $x_i$. In a recruitment phase, individuals modify their trait according recruitment signals send by successful individuals in the population. In particular, we assume that the intensity of the signal is proportional to the fitness of the recruiter. 

\begin{figure}[ht]
	\centering
	\includegraphics[width=\linewidth]{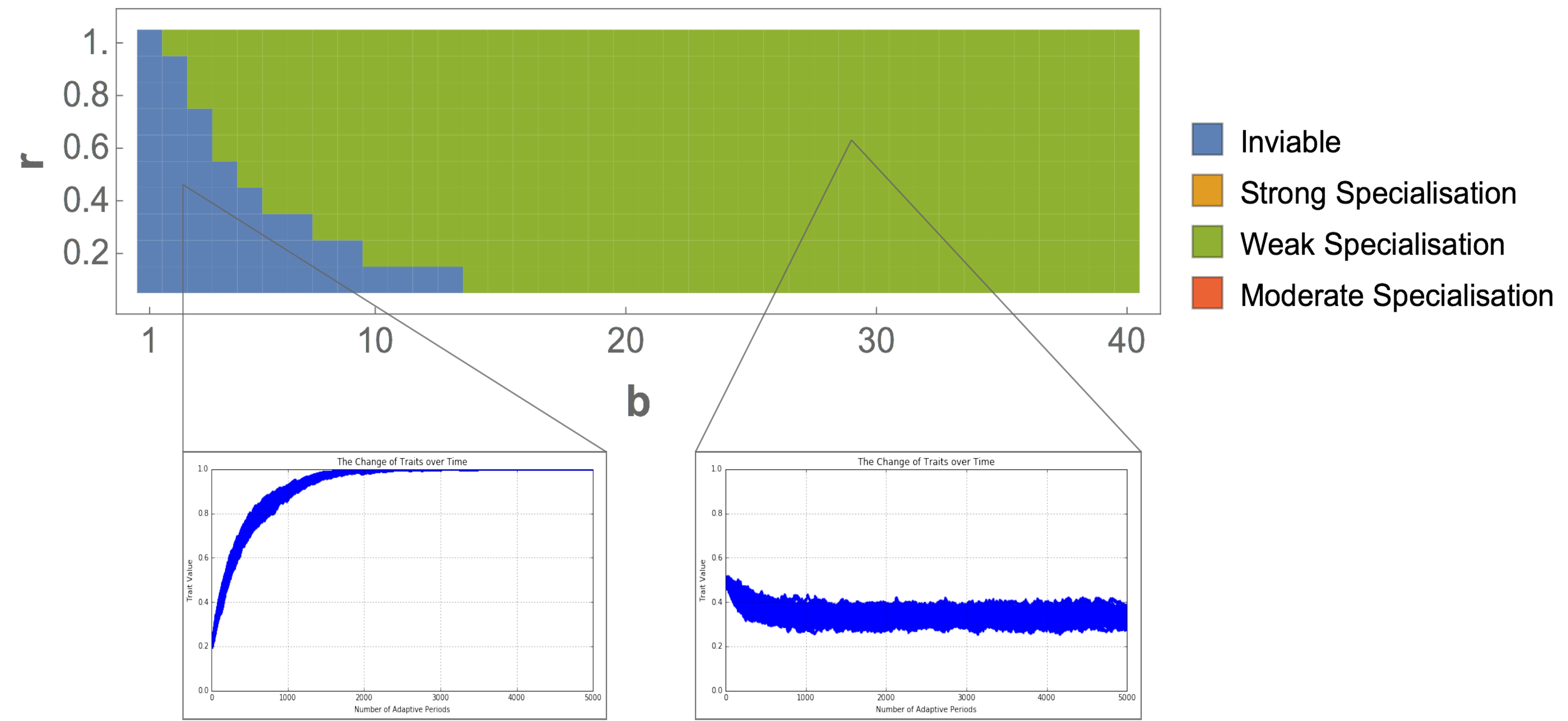}
    \caption{\label{fig:inset_recruitment} Classification of colonies based on the model of task recruitment}
\end{figure}

 The simulation results for this model are shown in Figure~\ref{fig:inset_recruitment}. Task recruitment favours weak specialisation across a large range of parameters.

\section{Discussion}

This paper shows that EGT is a promising avenue to study the effect of social interactions in task allocation models. In particular, we show that a simple model of social learning can give insights into when to expect different levels of specialization. We show that the mechanism by which individuals learn can dramatically change predictions: while social learning is conducive to societies of specialists and generalists, individual learning readily leads to strong specialization. 

More research is needed to understand how different species may rely on different kinds of learning. Our models suggest that the ecology of the tasks interacts in a non-trivial way with the cognitive capacities of the species.

Extensions of this model can introduce different types of tasks as well as the option of performing no tasks, which has been widely observed in social insect colonies. Further research is also needed to provide analytic predictions for individual learning, and to understand how different time-scales of learning and ecology may interact with each other in task allocation games.

\bibliographystyle{apalike}
\bibliography{social_learning.bib}

\end{document}